      \documentstyle[12pt,graphicx]{article}                
\begin{document}
\baselineskip 18pt
\def\today{\ifcase\month\or
 January\or February\or March\or April\or May\or June\or
 July\or August\or September\or October\or November\or December\fi
 \space\number\day, \number\year}
\def\thebibliography#1{\section*{References\markboth
 {References}{References}}\list
 {[\arabic{enumi}]}{\settowidth\labelwidth{[#1]}
 \leftmargin\labelwidth
 \advance\leftmargin\labelsep
 \usecounter{enumi}}
 \def\newblock{\hskip .11em plus .33em minus .07em}
 \sloppy
 \sfcode`\.=1000\relax}
\let\endthebibliography=\endlist
\def\lsim{\ ^<\llap{$_\sim$}\ }
\def\gsim{\ ^>\llap{$_\sim$}\ }
\def\r2{\sqrt 2}
\def\beq{\begin{equation}}
\def\eeq{\end{equation}}
\def\beqn{\begin{eqnarray}}
\def\eeqn{\end{eqnarray}}
\def\rmuu{\gamma^{\mu}}
\def\rmud{\gamma_{\mu}}
\def\PL{{1-\gamma_5\over 2}}
\def\PR{{1+\gamma_5\over 2}}
\def\sinW2{\sin^2\theta_W}
\def\AEM{\alpha_{EM}}
\def\mul{M_{\tilde{u} L}^2}
\def\mur{M_{\tilde{u} R}^2}
\def\mdl{M_{\tilde{d} L}^2}
\def\mdr{M_{\tilde{d} R}^2}
\def\mz2{M_{z}^2}
\def\c2b{\cos 2\beta}
\def\au{A_u}
\def\ad{A_d}
\def\cob{\cot \beta}
\def\v#1{v_#1}
\def\tb{\tan\beta}
\def\epem{$e^+e^-$}
\def\KK{$K^0$-$\bar{K^0}$}
\def\wi{\omega_i}
\def\xj{\chi_j}
\def\Wmu{W_\mu}
\def\Wnu{W_\nu}
\def\m#1{{\tilde m}_#1}
\def\mH{m_H}
\def\mw#1{{\tilde m}_{\omega #1}}
\def\mx#1{{\tilde m}_{\chi^{0}_#1}}
\def\mc#1{{\tilde m}_{\chi^{+}_#1}}
\def\mwi{{\tilde m}_{\omega i}}
\def\mxi{{\tilde m}_{\chi^{0}_i}}
\def\mci{{\tilde m}_{\chi^{+}_i}}
\def\mz{M_z}
\def\sw{\sin\theta_W}
\def\cw{\cos\theta_W}
\def\cb{\cos\beta}
\def\sb{\sin\beta}
\def\rwi{r_{\omega i}}
\def\rxj{r_{\chi j}}
\def\rfp{r_f'}
\def\Kik{K_{ik}}
\def\Fq2{F_{2}(q^2)}
\def\tw{\tan\theta_W}
\def\sec2w{sec^2\theta_W}

\begin{titlepage}

\  \
\vskip 0.5 true cm
\begin{center}
{\large {\bf Effects of large CP violating phases 
on $g_{\mu}-2$ in MSSM}}\\

\vskip 0.5 true cm
\vspace{2cm}
\renewcommand{\thefootnote}
{\fnsymbol{footnote}}
 Tarek Ibrahim$^a$ and Pran Nath$^b$  
\vskip 0.5 true cm
\end{center}
\noindent
{a. Department of  Physics, Faculty of Science,
University of Alexandria,}\\
{ Alexandria, Egypt}\\ 
{b. Department of Physics, Northeastern University,
Boston, MA 02115-5000, USA } \\

\vskip 1.0 true cm

\centerline{\bf Abstract}
\medskip
Effects of CP violation on the supersymmetric electro-weak correction
to the anomalous magnetic moment of the muon are investigated
with the most general allowed set of CP violating phases in MSSM. The
analysis includes contributions from the chargino and the neutralino
exchanges to the muon anomaly. The supersymmetric contributions
depend only on specific combinations of CP phases.
The independent set of such phases is classified. We analyse the
effects of the phases under the EDM constraints and show
that large CP violating phases can drastically affect the magnitude 
of the supersymmetric electro-weak contribution to $a_{\mu}$ and 
may even affect its overall sign.
\end{titlepage}

\section{Introduction}
As is well known supersymmetric theories contain many new sources
of CP violation which mostly arise from the phases of the soft 
SUSY breaking parameters and that such phases contribute to
the electric dipole moments (EMDs) of the electron and of the neutron.
Experimentally the electron and the neutron EDMs have very strict 
limits, i.e., for the neutron the limit is\cite{harris} 

\beq
 |d_n|< 6.3\times 10^{-26} ecm
 \eeq
 and for the electron the limit is\cite{commins}
 \beq
 |d_e|< 4.3\times 10^{-27} ecm
\eeq
and these limits impose stringent constrains on particle physics
models.
In SUSY/string models one normally expects CP violating phases O(1) 
and phases of this size typically lead to EDM predictions in such
models already in excess of the current experimental limits.
Of the possible remedies to this problem the conventional approach
has been to assume that the phases are small\cite{ellis,wein}, 
typically, $O(10^{2-3})$,
which, however, constitutes a fine tuning. Another possibility
suggested is to assume that the SUSY spectrum is heavy in the
several TeV region\cite{na}.
 Generally,  a heavy spectrum may constitutes
 fine tuning\cite{lung} except in certain limited domains of the 
parameter space\cite{lung1,moroi}.  
Further, such a heavy spectrum may lie 
outside the reach of even the Large Hadron Collider (LHC) and
thus  a disappointing scenario from the point of view
of particle physicists. A third more encouraging possibility
is that the large phases could indeed be there, but one 
escapes the experimental EDM constraints because of cancellations
among the various contributions to the EDMs. This possibility was
proposed in Ref.\cite{in1} and there have been 
 further verification and 
 developments\cite{in2,fo,bgk,bartl,prs} 
and applications such as in dark matter\cite{ffo,cin,ks,ffo1,gf,choi}, 
in  low energy processes\cite{kane,pw,barger,more},
and on other SUSY 
phenomena\cite{pilaftsis,demir,babu,bk,everett,accomando}. 

 The cancellation mechanism opens a new window on the SUSY parameter
 space where  large CP phases along with a light SUSY spectrum 
 can co-exist. Thus significant effects on SUSY phenomena can
 result. One of the quantities affected by CP phases is 
$a_{\mu}=(g_{\mu}-2)/2$, where $g_{\mu}-2$ is the anomalous
magnetic moment of the muon. 
This quantity is of considerable current interest since the new
Brookhaven experiment \cite{bnl1} will measure $a_{\mu}$ to 
an accuracy of better than a factor of 20\cite{bnl2}. 
Further, recently there has been considerable progress in reducing 
the hadronic error\cite{kino,davier,hayakawa}. With the reduced 
hadronic error the new $g_{\mu}$ experiment will test the
Standard Model (SM) electro-weak correction\cite{fuji} which including
the two loop SM corrections   
stands at\cite{czar}
\beq
 a_{\mu}^{SM}=15.1\times 10^{-10}
 \eeq
It turn out that the supersymmetric electro-weak corrections to
$a_{\mu}$ can be quite large and these 
 supersymmetric effects on $a_{\mu}$ have been investigated
for many years\cite{yuan,kos,lopez,chatto}. However, the CP violating 
supersymmetric electro-weak effects on  $a_{\mu}$ have been ignored  
for the 
reason that small CP phases or large CP phases with a heavy spectrum 
lead only to negligible effects on $a_{\mu}$.

 With the
cancellation mechanism the possibility of large CP phases along
with a light spectrum arises and such a situation can lead to
very significant effects on $a_{\mu}$. Indeed in a recent work\cite{in3}
the effects of CP phases on $a_{\mu}$ in the context of  
the minimal supergravity model (mSUGRA) were analysed and it was
shown that   CP violating phases can produce significant effects 
on  $a_{\mu}$. 
 In the absence of CP violating phases the 
soft SUSY breaking parameters at the GUT scale in
 mSUGRA\cite{chams} consist of 
the universal scalar mass $m_0$, the universal gaugino mass 
$m_{\frac{1}{2}}$, the universal trilinear coupling $A_0$ and
 $\tan\beta= <H_2>/<H_1>$ where $H_2$ gives  mass to the up quark
 and $H_1$ gives  mass to the down quark. 
 More generally the soft SUSY breaking parameters as well
 as the Higgs VEVs are complex and have phases. However, by 
 a redefinition of 
 fields it is easily seen that there are only two CP violating
 phases in mSUGRA. These can be chosen to be the phase of $A_0$
 and the phase of $\mu_0$ where $\mu_0$ appears in the Higgs 
 mixing term, i.e., in the term $\mu_0 H_1H_2$ in the superpotential.
 In this paper we
extend our analysis of the effects of CP violating phases on $a_{\mu}$
to supergravity models with 
non-universalities\cite{soni,kap,mata,ole,polo,nonuni,post,rev} 
and to the Minimal 
Supersymmetric Standard Model (MSSM) which has many more CP 
violating phases. 
The existence of a larger set of CP phases widens the region
of the parameter space where cancellations can occur. The purpose of
this paper is to derive the general one loop supersymmetric correction
to $a_{\mu}$  with the most general set of CP violating phases 
allowed in MSSM and determine the numerical effects of these CP violating 
phases  on $a_{\mu}$ under the experimental constraints 
on the electron and on the neutron EDM.

The outline of the rest of the paper is as follows: In Sec.2 we 
 derive the general one loop formula 
for $a_{f}$ for the case of a  fermion $f$ interacting
with a  fermion and a  scalar in the presence of 
CP violating phases and without any approximation on the relative
size of the external and internal particle masses in the loop. In Sec.3 we 
apply this formula for the computation of the chargino and
neutralino exchange contributions to $a_{\mu}$ for the most general
allowed set of CP violating phases in this sector.
In Sec.4 and in Appendix A we study the combination of CP phases that enter
$a_{\mu}$  and compare them with the corresponding combinations 
that arise in the expressions for the electron and the neutron EDMs.
In Sec.5 and in Appendix B the supersymmetric limit of our result
is given and it is explicitly shown how the one loop Standard Model
contribution to $a_{\mu}$ including the one loop QED 
correction,
i.e.  $\alpha_{em}/2\pi$, 
 is cancelled  by the supersymmetric contribution.
 In Sec.6 we give a discussion of the satisfaction of the EDM constraints.
In Sec.7 we give an analysis of CP violating effects on $a_{\mu}$.
 Conclusions are
given in Sec.8. Appendix C is devoted to a discussion of the 
vanishing CP violating phases  and a comparison of our results 
with previous analyses.
 
 \section{CP Effects on $g-2$ in MSSM}

We give here the general analysis for the CP effects on $g-2$ of a
fermion.  In general for the interaction of a  
fermion $\psi_f$ of mass $m_f$ interacting with a 
 fermion $\psi_i$ of mass $m_i$ and a  scalar $\phi_k$
 of mass $m_k$, the 
 vertex interaction  has the general form
\beq
-{\cal L}_{int}= \sum_{ik}\bar{\psi_f}(K_{ik}\PL +L_{ik}\PR)
\psi_i \phi_k + H.c. 
\eeq
This interaction violates CP invariance 
 iff $ Im(K_{ik}L_{ik}^*)\neq  0 $. The one loop contribution to 
 $a_f$   is given by
\beq
a_f=a_f^1+a_f^2
\eeq
where $a_f^1$ and $a_f^2$ arise from Fig. 1(a)
 and Fig. 1(b) respectively. 
$a_f^1$ is a sum of two terms:  $a_f^1=a_f^{11}+a_f^{12}$ where
\beq
a_f^{11}=\sum_{ik} \frac{m_f}{8 \pi^2 m_i}Re(K_{ik}L_{ik}^*)
I_1(\frac{m_f^2}{m_i^2},\frac{m_k^2}{m_i^2})
\eeq
and 
\beq
I_1(\alpha,\beta)=-\int_0^1dx\int_0^{1-x}dz 
\frac{z}{\alpha z^2+(1-\alpha-\beta)z+\beta}
\eeq
and where 
\beq
 a_f^{12}=\sum_{ik} \frac{m_f^2}{16 \pi^2 m_i^2}(|K_{ik}|^2+|L_{ik}|^2)
I_2(\frac{m_f^2}{m_i^2},\frac{m_k^2}{m_i^2})
\eeq
and
\beq
I_2(\alpha,\beta)=\int_0^1dx\int_0^{1-x}dz 
\frac{z^2-z}{\alpha z^2+(1-\alpha-\beta)z+\beta}
\eeq
Similarly, $a_f^2$  consists of two terms:
 $a_f^2=a_f^{21}+a_f^{22}$ where 
\beq
a_f^{21}=\sum_{ik}\frac{m_f}{8 \pi^2 m_i}Re(K_{ik}L_{ik}^*)
I_3(\frac{m_f^2}{m_i^2},\frac{m_k^2}{m_i^2})
\eeq
 and 
\beq
I_3(\alpha,\beta)=\int_0^1dx\int_0^{1-x}dz 
\frac{1-z}{\alpha  z^2+(\beta-\alpha -1)z+1}
\eeq
and where
\beq
a_f^{22}=-\sum_{ik}\frac{m_f^2}{16\pi^2 m_i^2}
(|K_{ik}|^2+|L_{ik}|^2)I_4(\frac{m_f^2}{m_i^2},\frac{m_k^2}{m_i^2})
\eeq
and

\beq
I_4(\alpha,\beta)=\int_0^1dx\int_0^{1-x}dz 
\frac{z^2-z}{\alpha  z^2+(\beta-\alpha -1)z+1}
\eeq
In the above we have given the exact expressions for the integrals 
$I_1-I_4$ rather than their approximate forms in the limit when 
one neglects terms of size $m_f^2$ relative to $m_k^2$ and $m_i^2$ which allows
one to write simple  closed form expressions for them. We will see
that the general expressions are needed to discuss the 
supersymmetric limit of our results which provides the absolute 
check on our normalizations.

\noindent
\section{$g_{\mu}-2$ with CP Violating Phases}

We apply now the above relations for the computation of the 
chargino and the neutralino exchange contributions. We consider
the chargino exchange contributions first.  The CP violating
phases enter here via the chargino mass matrix defined by

\beq
M_C=\left(\matrix{|\m2|e^{i\xi_2} & \r2 m_W  \sb e^{-i\chi_2}\cr
	\r2 m_W \cb e^{-i\chi_1}& |\mu| e^{i\theta_{\mu}}}
            \right)
\eeq
where $\chi_1$ and $\chi_2$ are phases of the Higgs VEVs, i.e.,
$<H_i>=|<H_i>| e^{i\chi_i}$ (i=1,2). 
The matrix of Eq.(14) can be diagonalized by the biunitary transformation
$U^* M_C V^{-1}=diag(\mc1,\mc2)$ where U and V are  unitary 
matrices.
By looking at the muon-chargino-sneutrino interaction one can identify
$K_{i}$ and $L_{i}$ and one finds 
\beq
a^{\chi^{-}}_{\mu}=a^{21}_{\mu}+a^{22}_{\mu}
\eeq
where $a^{21}_{\mu}$ and $a^{22}_{\mu}$ are given below. 
We exhibit these  only in the limit where $I_3(\alpha,\beta)$ and 
$I_4(\alpha,\beta)$ have their first arguments zero and one may write
\beq 
I_3(0,x)= -\frac{1}{2}F_3(x),~I_4(0,x)=-\frac{1}{6}F_4(x)
\eeq
where
\beq
F_3(x)=\frac{1}{(x-1)^3}(3x^2-4x+1-2x^2 lnx)
\eeq

\beq
F_4(x)=\frac{1}{(x-1)^4}(2x^3+3x^2-6x+1-6x^2 lnx).
\eeq
In the above approximation we have
\beq
a^{21}_{\mu}=\frac{m_{\mu}\alpha_{EM}}{4\pi\sin^2\theta_W}
\sum_{i=1}^{2}\frac{1}{M_{\chi_i^+}}Re(\kappa_{\mu} U^*_{i2}V^*_{i1})
F_3(\frac{M^2_{\tilde{\nu}}}{M^2_{\chi_i^+}}).
\eeq
and
\beq
a^{22}_{\mu}=\frac{m^2_{\mu}\alpha_{EM}}{24\pi\sin^2\theta_W}
\sum_{i=1}^{2}\frac{1}{M^2_{\chi_i^+}}
(|\kappa_{\mu} U^{*}_{i2}|^2+|V_{i1}|^2)
F_4(\frac{M^2_{\tilde{\nu}}}{M^2_{\chi_i^+}}).
\eeq
where
\beq
 \kappa_{\mu}=\frac{m_{\mu}}{\r2 M_W \cos\beta}e^{-i\chi_1}
 \eeq

\noindent

Next we discuss the neutralino exchange contribution to $a_{\mu}$.
CP violating effects here are all contained in the neutralino 
and smuon mass matrices. For the neutralino mass matrix the 
CP violating phases enter as below  
\beq
\left(\matrix{|\m1|e^{i\xi_1}
 & 0 & -\mz\sw\cb e^{-i\chi_1} & \mz\sw\sb e^{-i\chi_2} \cr
  0  & |\m2| e^{i\xi_2} & \mz\cw\cb e^{-i \chi_1}& -\mz\cw\sb e^{-i\chi_2} \cr
-\mz\sw\cb e^{-i \chi_1} & \mz\cw\cb e^{-i\chi_2} & 0 &
 -|\mu| e^{i\theta_{\mu}}\cr
\mz\sw\sb e^{-i \chi_1} & -\mz\cw\sb e^{-i \chi_2} 
& -|\mu| e^{i\theta_{\mu}} & 0}
			\right).
\eeq
 The neutralino mass  matrix $M_{\chi^0}$ is a 
 complex non hermitian and symmetric matrix and   
can be  diagonalized  using a unitary matrix X 
such that
$X^T M_{\chi^0} X$=${\rm diag}(\mx1, \mx2, \mx3, \mx4)$.
Since the loop correction involving the neutralino exchange also
involves the smuon exchange (see Fig.1a) the CP phases in the
 smuon $(mass)^2$ also enter the analysis. 
	The smuon $(mass)^2$ matrix is given by   
\beq
M_{\tilde{\mu}}^2=\left(\matrix{M_{\tilde{\mu}11}^2 & 
m_{\mu}(A_{\mu}^{*}m_0-\mu \tan\beta e^{i(\chi_1+\chi_2)})  \cr
	m_{\mu}(A_{\mu} m_0-\mu^{*} \tan\beta e^{-i(\chi_1+\chi_2)})
                            & M_{\tilde{\mu}22}^2}
				\right),
\eeq
 This matrix is  hermitian and can be diagonalized by the unitary transformation
\beq
D^\dagger M_{\tilde{\mu}}^2 D={\rm diag}(M_{\tilde{\mu}1}^2,
              M_{\tilde{\mu}2}^2)
\eeq
The neutralino exchange contribution to $a_{\mu}$ is given by
\beq
a^{\chi^0}_{\mu}=a^{11}_{\mu}+a^{12}_{\mu}
\eeq
where
\beq
a^{11}_{\mu}=\frac{m_{\mu}\alpha_{EM}}{2\pi\sin^2\theta_W}
\sum_{j=1}^{4}\sum_{k=1}^{2}\frac{1}{M_{\chi_j^0}} Re(\eta^k_{\mu j})
I_1(\frac{m_{\mu}^2}{M^2_{\chi_j^0}}, 
\frac{M^2_{\tilde{\mu_k}}}{M^2_{\chi_j^0}})
\eeq
and
\beqn
\eta^k_{\mu j} & &=-
(\frac{1}{\sqrt 2}[\tan\theta_W X_{1j}+ X_{2j}]D_{1k}^*
-\kappa_{\mu} X_{3j} D_{2k}^*)\nonumber\\
&&
(\sqrt 2\tan\theta_W X_{1j}D_{2k} +\kappa_{\mu} X_{3j} D_{1k})
\eeqn
and $a^{12}_{\mu}$ is given by 
\beq
a^{12}_{\mu}=\frac{m^2_{\mu}\alpha_{EM}}{4\pi\sin^2\theta_W}
\sum_{j=1}^{4}\sum_{k=1}^{2}\frac{1}{M^2_{\chi_j^0}}X^k_{\mu j}
I_2(\frac{m_{\mu}^2}{M^2_{\chi_j^0}}, 
\frac{M^2_{\tilde{\mu_k}}}{M^2_{\chi_j^0}})
\eeq
where
 
\beqn
X^k_{\mu j}&&=\frac{m^2_{\mu}}{2 M^2_W \cos^2\beta}|X_{3j}|^2 \nonumber\\
&&
+\frac{1}{2}\tan^2\theta_W |X_{1j}|^2
(|D_{1k}|^2+4|D_{2k}|^2)
+\frac{1}{2} |X_{2j}|^2|D_{1k}|^2\nonumber\\
&&
+\tan\theta_W |D_{1k}|^2 Re(X_{1j}X_{2j}^*)\nonumber\\
&&
+\frac{m_{\mu}\tan\theta_W}{M_W \cos\beta}
Re(e^{-i\chi_1}X_{3j}X_{1j}^*D_{1k}D_{2k}^*)\nonumber\\
&&
-\frac{m_{\mu}}{M_W\cos\beta}Re(e^{-i\chi_1}X_{3j}X_{2j}^*D_{1k}D_{2k}^*)
\eeqn
If one ignores
 the muon mass with respect to the other masses
involved in the problem, the form factors $I_1(\alpha,\beta)$
and $I_2(\alpha, \beta)$ become
\beq
I_1(0,x)=\frac{1}{2}F_1(x), I_2(0,x)=\frac{1}{6}F_2(x)
\eeq
where
\beq
F_1(x)=\frac{1}{(x-1)^3}(1-x^2+2x lnx)
\eeq
and 
\beq
F_2(x)=\frac{1}{(x-1)^4}(-x^3+6x^2-3x-2-6x lnx).
\eeq

\section{The number of independent linear combinations of phases
that enter $a_{\mu}$}
Not all the phases that enter in the chargino, neutralino and smuon
mass matrices are independent. We discuss here 
the set of independent phases that enter $a_{\mu}$. 
We consider the chargino contribution to $a_{\mu}$ first.  Here the
matrix elements of $U$ and $V$ 
as defined in the paragraph following Eq.(14) 
along with $\kappa_{\mu}$ as defined by Eq.(21) 
carry the phases
 $\xi_2$, $\theta_{\mu}$, $\chi_1$ and $\chi_2$.
 By introducing the transformation
$ M_C=B_R M_C'B_L^{\dagger}$ and choosing 
$B_R=diag(e^{i\xi_2},e^{-i\chi_1})$ and
$B_L=diag(1,e^{i(\chi_2+\xi_2)})$ we can rotate the phases so that
$M_C'$  is given by 

\beq
M_C'=\left(\matrix{|\m2| & \r2 m_W  \sb \cr
        \r2 m_W \cb & |\mu| e^{i(\theta_{\mu}+\xi_2+\chi_1+\chi_2)}}
            \right)
\eeq
The matrix $M'_C$  can be diagonalized by the biunitary transformation
$U_R^{\dagger}M_C'U_L$=diag
$(\mc1, \mc2)$.
It is clear that the matrix elements of $U_L$ and $U_R$  are functions only
of the combination $\theta=\theta_{\mu}+\xi_2+\chi_1+\chi_2$.
 We also have
$U^* M_C V^{-1}=diag(\mc1,\mc2)$ where
$U=(B_R U_R)^T$, and
V=$(B_LU_L)^\dagger$.  By inserting the new forms of $U$ and $V$ in the 
chargino contribution one finds
(as shown in Appendix A)
 that $a^{21}_{\mu}$
and  $a^{22}_{\mu}$ depend on only one combination, i.e.,  
$\theta = \theta_{\mu} + \xi_2 +\chi_1 +\chi_2$.

Now we turn to the neutralino contribution, the phases that enter here 
are $\theta_{\mu}$, $\alpha_{A_{\mu}}$, 
$\xi_2$, $\xi_1$, $\chi_1$ and $\chi_2$ and they
are carried by the matrix elements of $X$, $D_{\mu}$ and the phase of 
$\kappa_{\mu}$. 
Next we  make the transformation
$M_{\chi^0}$=$P_{\chi^0}^T$ $M_{\chi^0}'$ $P_{\chi^0}$
where
\beq
P_{\chi^0}=diag(e^{i\frac{\xi_1}{2}},e^{i\frac{\xi_2}{2}},e^{-i
(\frac{\xi_1}{2}
+\chi_1)},
e^{-i(\frac{\xi_2}{2}+\chi_2)})
\eeq
After the transformation the matrix $M_{\chi^0}^{'}$ takes the form
\beq
\left(\matrix{|\m1| & 0 & -\mz\sw\cb & \mz\sw\sb
                  e^{-i\frac{\Delta \xi}{2}} \cr
    0  & |\m2| & \mz\cw\cb e^{i\frac{\Delta \xi}{2}} & -\mz\cw\sb \cr
-\mz\sw\cb & \mz\cw\cb e^{i\frac{\Delta \xi}{2}} & 0 & -|\mu|e^{i\theta '} \cr
 \mz\sw\sb e^{-i\frac{\Delta \xi}{2}}  & -\mz\cw\sb & -|\mu|e^{i\theta '} & 0}
                        \right).
\eeq
where $\theta^{'} =\frac{\xi_1+\xi_2}{2}+\theta_{\mu}+\chi_1+\chi_2$,
and  $\Delta \xi=(\xi_1-\xi_2)$.
 The matrix $M_{\chi^0}^{'}$
can be diagonalized by the transformation
$Y^T M_{\chi^0}' Y$=${\rm diag}(\mx1, \mx2, \mx3, \mx4)$ where 
 Y is a function only of
$\theta^{'}$ and $\Delta \xi/2$. Thus the
 complex non hermitian and symmetric matrix $M_{\chi^0}$
can be  diagonalized  using a unitary matrix $X=P_{\chi^0}^{\dagger}Y$
such that
$X^T M_{\chi^0} X={\rm diag}(\mx1, \mx2, \mx3, \mx4)$.
As shown in Appendix A by applying the above transformations to each term of
$\eta^{k}_{\mu j}$ and $X^{k}_{\mu j}$ one finds that the phase 
 combinations that
enter here are $\theta^{'}$, $\Delta \xi/2$ and $\alpha_{A_{\mu}}+
\theta_{\mu}+\chi_1+\chi_2$ from
which we can construct the three combinations $\xi_1+\theta_{\mu}
+\chi_1+\chi_2$, $\xi_2+\theta_{\mu}+\chi_1+\chi_2$ and 
$\alpha_{A_{\mu}}+\theta_{\mu}+\chi_1+\chi_2$.

By defining $\theta_1=\theta_{\mu}+\chi_1+\chi_2$ one finds that
the $a_{\mu}$ dependence on phases from both the chargino and the
neutralino exchanges consists only  of the three combinations
$\alpha_{A_{\mu}}+\theta_1$, $\xi_1+\theta_1$ and $\xi_2+\theta_1$.
One may compare these combinations with those that appear in the 
supersymmetric contribution to the 
electron and the neutron EDMs. In the analysis of Ref.\cite{in2} we found 
 that
the electron and the neutron EDMs depend on the following combinations:
$\xi_i+\theta_1$, $(i=1,2,3)$ and $\alpha_{A_{k}} +\theta_1$ with
$k=u,d,t,b,c,s;\it l$.
We note that even though $a_{\mu}$ and the EDMs are very different
physical quantities the linear combination of phases that enter 
in them are similar. In fact  the phases that enter $a_{\mu}$
are a subset of phases that enter in the supersymmetric contributions
to the EDMs.

\section{The Supersymmetric Limit}
To check the absolute normalization of our results we 
 discuss now their supersymmetric limit.
In Ref.\cite{in3}  the supersymetric contributions to $a_{\mu}$ from the
chargino sector in  the supersymmetric limit were computed by
going to the limit such that   	
	
\beq	
	 U^*M_CV^{-1}=diag(M_W,M_W)
\eeq
In this limit it was shown that the contributions from this sector
was precisely negative of the contribution from the W exchange\cite{fuji}.
The analysis of Ref.\cite{in3} was carried out in the framework of 
mSUGRA with two CP violating phases. For the MSSM case being discussed
here with many CP phases 
 the structure of $U$ and $V$ matrices
in susy limit will be modified so that 
\beq
 U=\frac{1}{\sqrt 2}\left(\matrix{1 & e^{-i\chi_1} \cr
        -1&  e^{-i\chi_1}}
            \right),
 V=\frac{1}{\sqrt 2}\left(\matrix{1 &  e^{-i\chi_2} \cr
        1& - e^{-i\chi_2}}
            \right)
\eeq
However, taking  the phases $\chi_1$ and $\chi_2$ 
into account we find exactly the same result as in Ref.\cite{in3}
due to the appearance of the $\kappa_{\mu}$ phase.
Thus the sum of the  W exchange contribution and of the chargino 
exchange contributions cancel in the supersymmetric limit. 
Similarly it was shown in Ref. \cite{in3}
by making a unitary transformation that 
the neutralino mass matrix in the supersymmetric limit 
can be written in the form 
 
    \beq
  X^T M_{\chi^0} X= diag(0,0, M_Z, M_Z)
  \eeq
where the eigen-values are positive definite. It was then shown 
 that the last two eigen-modes give a  contribution which is 
negative of the contribution from the Z exchange in the 
Standard Model\cite{fuji}.
In the case of MSSM we are discussing here 
the structure of the diagonalizing matrix $X$ is now changed 
because of the $\chi_1$ and $\chi_2$ phases (see Appendix B). 
However, the final
result we arrived at in Ref.\cite{in3} still holds due to the 
appearance of the  
$\kappa_{\mu}$ phase once again.
Thus the sum of the Z exchange and of the heavy neutralino exchanges
exactly cancel in the supersymmetric limit. 

 We now turn to the supersymmetric limit of the contribution from the
first two massless eigen states of the neutralino mass matrix.
A direct sum over the first two eigen-modes for the case of Eq.(38) 
in the supersymmetric limit gives
 \beq
 a_{\mu}^{susy}(zero-modes)=-\frac{\alpha_{em}}{2\pi}
 \eeq
Thus the sum of the Standard Model contributions to $a_{\mu}$ from the
photon at one loop and form the Z and the W exchanges at one 
loop\cite{fuji} is 
cancelled by the supersymmetric 
contributions from the neutralino 
and the chargino exchanges in MSSM at one loop 
in the supersymmetric limit, i.e., in the supersymmetric limit one 
has 

\beq
a_{\mu}^{MSSM}=0
\eeq 
This result is consistent with the expectation on general 
grounds\cite{fr,bg}. 
 The details of the derivation of Eq.(39) are given in 
 Appendix B.
 
\section{Satisfaction of EDM Constraints} 	

Before proceeding to discuss the CP effects on $a_{\mu}$
we describe briefly the EDM constraints on the CP violating
phases. As is well known for the case of the neutron EDM there are 
three operators that contribute to the neutron EDM, namely,
the electric dipole moment operator, the color dipole moment operator
 and the purely gluonic dimension six operator. Both the electric and 
 color operators have three components each from the chargino, neutralino 
 and gluino contributions. For the electron case we have only
the electric dipole moment operator which has only two components, the
chargino and the neutralino ones. Recently, it has been pointed out
 that in addition to the above contributions certain two loop graphs may
 also contribute significantly in some
 regions of the parameter space\cite{darwin}. In our analysis here we 
 include the effects of these contributions as well.  However,
  the effect of these terms is found to be generally very small compared 
  to the other contributions in most of the parameter space we consider.
Satisfaction  of the EDM constraints can be achieved in a 
straightforward fashion using the cancellation mechanism.

\section{Analysis of CP Violating Effects}
In the above we have given the most general analysis of $a_{\mu}$ 
within the framework of MSSM with inclusion of CP violating
phases. Our results limit to those of Refs.\cite{yuan,kos} in the
limit when CP violating effects vanish (see Appendix C for details).
For the case of the general analysis with phases in MSSM 
the number of parameters that
enter $a_{\mu}$ along with the number of parameters that enter
the EDM constraints which must be imposed on the CP violating
phases is  large. For the purpose of a numerical study of the
CP violating effects on $a_{\mu}$ we shall confine ourselves to 
a more constrained set. Here we shall generate the masses of the
 sparticles at low energy starting with parameters at the GUT scale
and evolve these downwards using renormalization group equations.
At the GUT scale we shall use the parameters $m_0$, $m_{1/2}$, and
$A_0$, and $\mu$ will be determined via radiative breaking of the
electro-weak symmetry.
We set $\chi_1+\chi_2=0$  
and choose the phases that we vary to consist of  $\theta_{\mu}$,
 $\alpha_{A_0}$, and $\xi_i$ $(i=1,2,3)$. The choice of the above
 constrained set is 
 simply for the purpose of reducing the number of parameters for
 the numerical study.

We begin our discussion of the numerical results by exhibiting
the dependence of $a_{\mu}$ on the CP violating phases but without
the imposition of the EDM constraints. The dependence of $a_{\mu}$
on $\theta_{\mu}$ and $\alpha_{A_0}$ was already studied in 
Ref.\cite{in3} and we confine ourselves here to the dependence of
$a_{\mu}$ on $\xi_1$ and $\xi_2$. In Fig.2 we exhibit the
dependence of $a_{\mu}$ on $\xi_1$ and in Fig.3  the dependence
of $a_{\mu}$ on $\xi_2$. From Figs.2 and 3 we find that $a_{\mu}$
is significantly affected by the dependence of both 
$\xi_1$ and $\xi_2$.  However, a comparison of Fig.2 and Fig.3  
shows that the dependence of $a_{\mu}$ on $\xi_2$ is much
stronger than on $\xi_1$. The reason behind this  difference
is easily understood. The relatively weaker dependence on the $\xi_1$ 
phase arises because this phase appears only in the neutralino 
contribution while the $\xi_2$ phase appears both in the 
neutralino and in the chargino contributions to $a_{\mu}$.

 We discuss  now the effects of CP violating phases on $a_{\mu}$
under the EDM constraints.
In Table 1 we show four points which lie on the curves of
Fig.2 and Fig.3. As one can see the SUSY mass parameters
$m_0$ and $m_{1/2}$ are relatively small (i.e., $m_0,m_{1/2}<< 1~TeV$),
the CP phases are large and there is compatability with
experimental constraints on the electron EDM and on the neutron
EDM as a consequence of the cancellation mechanism.  
One also finds on comparing $a_{\mu}$ with and without phases
that the effects of CP violating phases on $a_{\mu}$ are very 
significant.

\begin{center} \begin{tabular}{|c|c|c|c|c|c|c|}
\multicolumn{6}{c}{Table~1:  } \\
\hline
  & $\theta_{\mu_0}$& $\alpha_{A_0}$ & $d_n(10^{-26} e cm)$&
  $d_e(10^{-27} e cm)$  & $ [a_{\mu}
(phases)](10^{-9})$
&$[a_{\mu}(0)](10^{-9})$\\
\hline
(1) &2.35 & $.4$ &$-3.08$ & $-0.86$ & $-4.8$& 7.45 \\
\hline
(2) & $1.98 $ & $0.4 $& $-0.34$ & $-1.67$ & $-7.8$ &$11.7$ \\
\hline
(3) & $1.2 $ & $-1.5 $& $ 1.87 $ & $2.24 $ & $-3.25$ & $5.6$ \\
\hline
(4) & $2.7 $ & $-0.4 $& $1.87 $ & $ -0.03 $ & $-15.5  $ & $3.15$\\
\hline

\end{tabular}\\
\noindent
Table caption: Parameters other than those 
exhibited corresponding to the cases (1)-(4)
are: (1) $m_0$=70, $m_{1/2}$=99, $tan\beta$=3 , $|A_0|$=5.6,
$\xi_1$=$-1$, $\xi_2$=1.5, $\xi_3$=0.62;
(2) $m_0$=80, $m_{1/2}$=99, $tan\beta$=5 , $|A_0|$=5.5,
$\xi_1$=$-0.8$, $\xi_2$=1.5, $\xi_3$=0.95;
(3) $m_0$=75, $m_{1/2}$=132, $tan\beta$=4 , $|A_0|$=6.6,
$\xi_1$=$-1$, $\xi_2$=1.78, $\xi_3$=2.74;
(4) $m_0$=70, $m_{1/2}$=99, $tan\beta$=6 , $|A_0|$=3.2,
$\xi_1$=0.63, $\xi_2$=0.41, $\xi_3$=0.47,
where all masses are in GeV units and all phases are in rad. 
\end{center}
 
 In Fig.4 we exhibit $a_{\mu}$ as a function of $m_{1/2}$ 
 where all points on these
 trajectories satisfy the experimental constraints on the
 electron EDM and on the neutron EDM by cancellation.
 One finds that the magnitude
 of the supersymmetric electro-weak 
  contributions are comparable to and even larger than
 the Standard Model electro-weak contribution as 
 given by Eq.(3)\cite{czar}.
   
\section{Conclusions}
 We have  given in this paper a complete
one loop analysis of the effects of CP violating phases on 
$a_{\mu}$ with the most general set of allowed phases in MSSM
in this sector. We have checked the absolute normalization of our results
exhibiting the complete cancellation of the supersymmetric result
in the supersymmetric limit with the Standard Model result
including the qed one loop correction to $a_{\mu}$, i.e.,
 $\alpha_{em}/2\pi$.  A detailed numerical
analysis of the CP violating effects on $a_{\mu}$ for 
the regions which satisfy the EDM constraints is also
given. 
Computations of $a_{\mu}$ 
under the EDM constraints shows that the supersymmetric electro-weak
effects can generate significant 
contributions to  $a_{\mu}$ even with moderate values of $\tan\beta$,
i.e., $\tan\beta \sim 3-6$, which can be comparable to the Standard 
Model electro-weak correction. Thus supersymmetric CP effects on
$a_{\mu}$ are within the realm of observability in the new
Brookhaven $g_{\mu}-2$ experiment.\\

\noindent
{\bf Acknowledgements}\\ 
 This research was supported in part by NSF grant 
PHY-9901057. \\
\noindent
{\bf Appendix A:}\\ 
In this appendix we give the explicit derivation of the linear
combinations of phases on which
$a_{\mu}$ depends.
For the chargino contributions the phases are contained in the 
quantities $Re(\kappa_{\mu} U^*_{i2}V^*_{i1})$
 and $|\kappa_{\mu} U^{*}_{i2}|^2+|V_{i1}|^2$.
 By using $U=(B_{R} U_{R})^T$ and $V=(B_{L}U_{L})^{+}$
 as defined in Sec.4, where
$U_L$ and $U_R$ are functions of the combination 
$\theta=\theta_{\mu}+\xi_2+\chi_1+\chi_2$, one finds
\beq
U^{*}_{i2} = e^{i \chi_1} U^{*}_{R2i}
\eeq
and 
\beq
V^{*}_{i1} = U_{L1i}
\eeq
which leads to 
\beq
\kappa_{\mu} U^*_{i2}V^*_{i1}  =  |\kappa_{\mu}|  U^{*}_{R2i}  U_{L1i}
\eeq
and 
\beq
|\kappa_{\mu} U^{*}_{i2}|^2+|V_{i1}|^2 = |\kappa_{\mu}|^2 |U^{*}_{R2i}|^2 +
|U_{L1i}|^2
\eeq
Eqs.(43) and (44) show that the chargino contribution to $a_{\mu}$ 
depends only  on one combination of phases, i.e.,
$\theta=\theta_{\mu}+\xi_2+\chi_1+\chi_2$.
  
For the case of the neutralino contribution to $a_{\mu}$,
 the phases are contained in the quantities
$Re(\eta^k_{\mu j})$ and $X^{k}_{\mu j}$. 
We first consider the quantity $Re(\eta^k_{\mu j})$.
 It consists of six terms in the product
\beqn
\eta^k_{\mu j} & &=
([a X_{1j}+ b X_{2j}]D_{1k}^*
-\kappa_{\mu} X_{3j} D_{2k}^*)\nonumber\\
&&
(c X_{1j}D_{2k} +\kappa_{\mu} X_{3j} D_{1k})
\eeqn
where $a$, $b$ and $c$ are real numbers and independent of phases. 
 The first term in the expansion of Eq.(45) is 
\beq
a c X^{2}_{1j} D_{1k}^*  D_{2k}=
\pm a c X^{2}_{1j} \cos \theta_f \sin \theta_f e^{i \beta_f}
\eeq
where $+(-)$ sign is for $k=1(2)$ and
where the following definitions are used
\beq
\tan 2\theta_f = \frac{2m_{\mu}[|m_{0} A_{\mu}|^2+|\mu R_{\mu}|^2-
2|m_{0} A_{\mu} \mu R_{\mu}|\cos \alpha]^{1/2}}
{M^2_{\tilde{\mu}11}-M^2_{\tilde{\mu}22}}
\eeq
Here $R_{\mu}=\tan\beta e^{i(\chi_1+\chi_2)}$ and $\alpha=
\alpha_{A_{\mu}}+\theta_{\mu}+\chi_1+\chi_2$.
The phase $\beta_f$ is defined such that
\beq
\cos\beta_f=\frac{A}{[A^2+B^2]^{1/2}}
\eeq
and 
\beq
\sin\beta_f=\frac{B}{[A^2+B^2]^{1/2}}
\eeq
where $A$ is defined by
\beq
A = |m_0 A_{\mu}| \cos\alpha_{A_{\mu}}
-|\mu R^{*}_{\mu}|\cos(\theta_{\mu}+\chi_1
+\chi_2)
\eeq
and $B$ is defined by
\beq
B = |m_0 A_{\mu}| \sin\alpha_{A_{\mu}}
+|\mu R^{*}_{\mu}|\sin(\theta_{\mu}+\chi_1
+\chi_2)
\eeq
By using $X_{1j}= Y_{1j} e^{-i\xi_1/2}$
 where $Y_{1j}$ are functions only of $\theta^{'}$ and $\Delta \xi/2$,
 one can write the
first term as given by Eq.(46) as follows
 
\beq
a c X^{2}_{1j} D_{1k}^*  D_{2k}=
a c Y^{2}_{1j} f_{k}(\alpha) e^{-i(\xi_1-\beta_f)}
\eeq
where $f_{k}(\alpha)$ are real functions of $\alpha$. 
By using the  definition of $\beta_f$ as given by Eqs.(48) and (49)
and by taking the real part of Eq.(52) we find that the 
 right hand side of Eq.(52) contains the
 three combinations
$\theta^{'}$, $\Delta \xi/2$ and $\alpha$  which come from the 
first part of the
 the right hand side of Eq.(52) and in addition it contains  
the following two combinations: $\alpha_{A_{\mu}}-\xi_1$ and
 $\theta_{\mu}+\chi_1+\chi_2+\xi_1$ which come from the exponent.
  But the latter two combinations are
linear combinations of the first three combinations.
Thus the left hand side of Eq.(46) or Eq.(52) will depend only on the
combinations $\theta^{'}$, $\Delta \xi/2$ and $\alpha$.
The same analysis can be applied to the other five terms and each one
of them will give us the same three linear combinations.

 Next we consider  $X^{k}_{\mu j}$.
 It consists of six terms and the quantities 
in them which contain phases are $|X_{3j}|^2$,
 $ |X_{1j}|^2(|D_{1k}|^2+4|D_{2k}|^2)$, 
$|X_{2j}|^2|D_{1k}|^2$, $|D_{1k}|^2 Re(X_{1j}X_{2j}^*)$,
$Re(e^{-i\chi_1}X_{3j}X_{1j}^*D_{1k}D_{2k}^*)$
and $Re(e^{-i\chi_1}X_{3j}X_{2j}^*D_{1k}D_{2k}^*)$.
The first one of them, i.e. $|X_{3j}|^2$,
 can be written in terms of the $Y$ matrix as
$|Y_{3j}|^2$ which depends only on the two combinations
$\theta^{'}$ and $\Delta \xi/2$.
The second expression can be written as
$|Y_{1j}|^2 g_{k}(\alpha)$
where $g_{k}(\alpha)$  are real functions of 
$\alpha$ as defined after Eq.(47). 
 So this term will depend on the three combinations
$\theta^{'}$, $\Delta \xi/2$ and $\alpha$.
The third expression is similar to the second one and will give the same 
combinations.
The fourth expression can be written as
$h_{k}(\alpha) Re(Y_{1j} Y^{*}_{2j})$
where $h_{k}(\alpha)$ are real functions of $\alpha$.
 So this term also depends on the same three combinations.
The fifth expression can be written as
\beq
Re(e^{-i\chi_1}X_{3j}X_{1j}^*D_{1k}D_{2k}^*)  =
Re(Y_{3j}Y^{*}_{1j} s_{k}(\alpha) e^{i(\xi_1-\beta_f)})
\eeq
where $s_{k}(\alpha)$  are real functions of $\alpha$.
 By treating the exponential
term as we did in the first term of $\eta^{k}_{\mu j}$, 
it will give us two extra combinations
besides the usual three. These  are
$\alpha_{A_{\mu}}-\xi_1$ and $\theta_{\mu}+\chi_1+\chi_2+\xi_1$
 which, however, are linear combinations of the usual three.
  Thus we 
end up here with the same three combinations.
The sixth expression can be written as
\beq
Re(e^{-i\chi_1}X_{3j}X_{2j}^*D_{1k}D_{2k}^*) =
Re(Y_{3j} Y^{*}_{2j} s_{k}(\alpha) e^{i (\frac{\xi_1+\xi_2}{2}-\beta_f)})
\eeq
from which we can identify the usual three combinations 
and in addition one has the following combinations in the exponent
 :$\alpha_{A_{\mu}}-\frac{\xi_1+\xi_2}{2}$ 
and $\theta_{\mu}+\chi_1+\chi_2+\frac{\xi_1+\xi_2}{2}$.
These again are linear combination of the usual three and thus we
end up with only three phases in the neutralino contribution, i.e.,
$\theta^{'}$, $\Delta \xi/2$ and $\alpha$.

\noindent
{\bf Appendix B }\\
In this appendix we discuss the supersymmetric limit in the
massless sector..
For this purpose  we begin by exhibiting 
 the unitary matrix X that diagonalizes the neutralino
mass matrix in the supersymmetric limit
 such that the eigen-values are arranged so that
\beq
X^T M_{\chi^0}X=diag(0,0,M_Z,M_Z)
\eeq
With the above ordering the 
unitary matrix X takes on the form
\beq
\left(\matrix{\alpha & \beta & \frac{\sin\theta_W}{\sqrt 2}
  &  i\frac{\sin\theta_W}{\sqrt 2}   \cr
  \alpha\tw & \beta\tw  & -\frac{\cos\theta_W}{\sqrt 2}
     & -i\frac{\cos\theta_W}{\sqrt 2}   \cr
 \alpha e^{i \chi_1}&  -\frac{1}{2}\beta \sec2w e^{i \chi_1}
 & -\frac{1}{2} e^{i \chi_1}& \frac{i}{2}e^{i \chi_1} \cr
  \alpha e^{i \chi_2}& -\frac{1}{2}\beta \sec2w e^{i \chi_2}& 
\frac{1}{2} e^{i \chi_2}& -\frac{i}{ 2}e^{i \chi_2} }
                        \right).
\eeq
where 
\beq
\alpha = \frac{1}{\sqrt{3+\tan^2\theta_W}}, 
\beta=\frac{1}{\sqrt{1+\tan^2\theta_W+\frac{1}{2}sec^4\theta_W}}
\eeq
Using these results it is easily seen that the sum over the first two
neutralino mass eigen-values gives

\beq
a^{11}_{\mu}(zero-modes)=-\frac{\alpha_{EM}}{2\pi\sin^2\theta_W}
H \sum_{j=1}^{2}\sum_{k=1}^{2} Re(\eta^k_{\mu j})
(\frac{M_{\chi_j^0}}{m_{\mu}})\rightarrow 0
\eeq
where we set $M_{\tilde{\mu}k}= m_{\mu}$ and the factor $H$ is defined by
\beq
H=\int_0^1dx\int_0^{1-x}dz
\frac{z}{(z-1)^2}
\eeq
Thus in the supersymmetric limit the entire supersymmetric contribution to 
$a_{\mu}^1$ from the masseless neutralino states comes from 
$a_{\mu}^{12}$.
To compute this contribution  we need the sum  
\beq
a^{12}_{\mu}=\frac{m^2_{\mu}\alpha_{EM}}{4\pi\sin^2\theta_W}
\sum_{j=1}^{2}\sum_{k=1}^{2}\frac{1}{M^2_{\chi_j^0}}X^k_{\mu j}
I_2(\frac{m_{\mu}^2}{M^2_{\chi_j^0}},
\frac{M^2_{\tilde{\mu_k}}}{M^2_{\chi_j^0}})
\eeq
where the sum over j runs only  over the first two modes.
In the supersymmetric limit we set $M^{2}_{\tilde{\mu k}}=m^{2}_{\mu}$,
 $M_{\chi_j^0}\rightarrow 0$ (j=1,2), 
and 
$x_{\mu j}\equiv \frac{m_{\mu}^2}{M^2_{\chi_j^0}}\rightarrow \infty$
 (j=1,2) and 
\beq
\frac{m_{\mu}^2}{M^2_{\chi_j^0}}I_2(x_{\mu j},x_{\mu j})
\rightarrow  -\frac{1}{2}
\eeq
Now substitution of the explicit form of the X matrix gives
\beq
\sum_{j=1}^{2}\sum_{k=1}^{2}X^k_{\mu j}=4 sin^2{\theta_W}
\eeq
Use of Eqs. (61) and (62) in Eq.(60) gives 
\beq 
a_{\mu}^{12}(zero-modes)=-\frac{\alpha_{em}}{2\pi}
\eeq
 Thus we find that in the supersymmetric limit the exchange 
 of two massless neutralinos  gives a one loop contribution
 to $a_{\mu}$ 
which is exactly negative of the photonic one loop contribution.
Thus in the supersymmetric limit the sum of the one loop contributions
of the zero modes of the theory cancel. The cancellation provides
an absolute check on the normalization of our supersymmetric result
in this sector.

\noindent
{\bf Appendix C}\\
In this section we consider the limit of vanishing CP violating
phases and compare our results with those of previous works.
 We first compare our results with those 
of Ref.\cite{yuan}.
We  consider the chargino contribution
first. Using Eq.(2.8) of Ref.\cite{yuan} and noting that the 
free part of the Lagrangian density for the complex scalar fields in 
that work is given by $\frac{1}{2}(\partial_{\mu}z^*\partial^{\mu}z
-m^2z^*z)$, we find that our $K_i$ and $L_i$ are related to the
$A_L^{\pm}$ and $A_R^{\pm}$ of Ref.\cite{yuan} as follows: 

\beq
K_{1,2 \nu}\rightarrow-i\r2 A^{+,-}_L, 
L_{1,2 \nu}\rightarrow-i\r2 A^{+,-}_R
\eeq
Further, our form factors $F_3(x)$ and $F_4(x)$ are related to the
form factors $F_1$ and $F_2$ of Ref.\cite{yuan} as follows:
\beq
F_3(x)=-F_2(x),~~ F_4(x)=F_1(x)
\eeq
Defining 
\beq
 g_{1}^{\tilde{W}} = 2 a^{22}_{\mu},~~g_{2}^{\tilde{W}} = 2 a^{21}_{\mu}
\eeq
we find that our Eq.(12) in the limit of vanishing CP violating 
phases is given in the notation of Ref.{\cite{yuan}} by
\beq
g_{1}^{\tilde{W}}=\frac{m_{\mu}^2}{24\pi^2}
\sum_{a=1,2}\frac{A_R^{(a)2}+A_L^{(a)2}}{\tilde m_a^2}F_1(x_a)
\eeq
and similarly our Eq.(10) in the same limit in the notation of
Ref.\cite{yuan} is given by 
\beq
g_{2}^{\tilde{W}}=\frac{m_{\mu}}{4\pi^2}
\sum_{a=1,2}\frac{A_R^{(a)}A_L^{(a)}}{\tilde m_a}F_2(x_a)
\eeq
where 
\beq 
x_a=\frac{\tilde m_{\nu}^2}{\tilde m_a^2};~~a=1,2
\eeq
Eqs. (67) and (68) agree precisely with Eqs.(2.6a) and (2.6b) 
of Ref.\cite{yuan}
to leading order in $\mu^2/\tilde m_a^2$ taking account of the typo
in Eq.(2.6a) where $A_R^{(a)}$ should read $A_R^{(a)2}$ and noting 
that $A^{2 +,-}_L$  is proportional to $m_{\mu}^2/M_W^2$ and thus
does not contribute to leading order. 

We consider next the  neutralino contribution. From  the interaction 
Lagrangian Eq.(2.4) of Ref.\cite{yuan} we find 
the transition from our notation to that of Ref.\cite{yuan}as follows:
\beq
K_{kr}\rightarrow-\r2 i(O^{'}_{1r}B^{L}_{k}-C_{k}O^{'}_{2r}),
L_{kr}\rightarrow-\r2 i(O^{'}_{2r}B^{R}_{k}+C_{k}O^{'}_{1r})
\eeq
where we identify $O'$ to be 
\beq
O^{'}=\left(\matrix{\cos\delta & \sin\delta\cr
        -\sin\delta& \cos\delta}
            \right)
\eeq
Noting that  our form factors $F_1(x)$ and $F_2(x)$ are related to the
form factors $G_{2}(x)$ and $G_{1}(x)$ of Ref.\cite{yuan} by
\beq
F_1(x)=-G_2(x),~~F_2(x)=-2G_1(x)
\eeq
and defining 
$g_{1}^{\tilde{Z}} = 2 a^{12}_{\mu}$ and
$g_{2}^{\tilde{Z}} = 2 a^{11}_{\mu}$ 
we find that our Eq.(8) gives precisely Eq.(2.10) of Ref.\cite{yuan} 
taking account of the typos in Eq.(2.10a) in that $1/\tilde \mu_k$
should read $1/\tilde \mu_k^2$ and $G_2(x_{2k})$ in the same
equation should
read $G_1(x_{2k})$. Further, our Eq.(6) agrees precisely with 
Eq.(2.12) of Ref.\cite{yuan}.

 Next we  compare our results with those of Ref.\cite{kos}. 
For this purpose in the chargino sector we identify  $\tilde W_1$ 
and $\tilde W_2$ states with the states $\tilde W^-$ and $\tilde H^-$ of 
Ref.\cite{kos} in order to use Table 1 of Ref.\cite{kos}. 
With this identification  in the
limit of vanishing CP violating phases we find that in the
chargino sector our matrices V and U are real and orthogonal and 
are related to the matrices $O_1$ and $O_2$ of Ref.\cite{kos}
as follows

\beq
V_{km}^*\rightarrow O_{1mk},~~
U_{km}^*\rightarrow O_{2mk}
\eeq 
The analysis of Ref.{\cite{kos}} computes only the contribution 
 $a^{21}_{\mu}$ of $a^{\chi^{-}}_{\mu}$
 in their Eq.(5). Relating our $F_3(\eta)$ to
 their $F_{s\nu}(\eta)$ by $F_3(\eta)=-2F_{s\nu}(x)$, 
 we find that our Eq.(19) can be written in the
form 

\beq
2a^{21}_{\mu}=-\frac{m_{\mu} e^2}{4\pi^2 \sin^2\theta^2_W}\sum_{k}
\frac{m_{\mu}}{\r2 M_k m_W \cos\beta} O_{2 2k}O^T_{1 k1} F_{s\nu}
(\eta^{'}_{\nu k})
\eeq
which is exactly Eq.(5) of Ref.\cite{kos} on relating their $\sin\theta_H$
to our $\cos\beta$ by  $\sin\theta_H=\cos\beta$. 
The consistency of the analysis of Ref.\cite{kos} with our analysis,
however, requires that the sign of the terms with $M_W$ in
the chargino mass matrix given by Eq.(3b) of Ref.\cite{kos} be
reversed.

 To compare our results to those of Ref.\cite{kos} in the 
neutralino sector we note that our $H^{0}_1$ and $H^{0}_2$ states are
related to their $H^0$ and $H^{0'}$ states by $H^{0}_{1} = H^{0}$ and 
$H^{0}_{2} = H^{0'}$.  In the limit of vanishing CP violating phases,
our neutralino and smuon mass matrices become real, and the 
corresponding diagonalizing matrices X and D become orthogonal and
can be identified with the real orthogonal matrices O and S of 
Ref.\cite{kos}:
\beq
X\rightarrow O,~~ D\rightarrow S
\eeq 
The consistency of the analysis of Ref.\cite{kos} with our analysis,
however, requires that the sign of the terms with $M_Z$ in 
the neutranlino mass matrix given by Eq.(3a) of Ref.\cite{kos} be
reversed. 
The analysis of Ref.\cite{kos} calculated only the part $a^{11}_{\mu}$ 
in their Eq.(6). To compare the result of $a^{11}_{\mu}$
of our Eq.(25) with 
their Eq.(6) we first note that our $F_1$ is related to their F by 
$F_1(\eta)=-F(\eta)$. Second we need to identify the fields $\tilde W_i$
(i-1,2,3) in Eq.(6) of Ref.\cite{kos} in order to use Table 1 
of Ref.\cite{kos} to
 write out in detail the interactions of Eq.(6). This identification 
 is as follows:

\beq
\tilde{W}_1=\tilde{B}^0, 
\tilde{W}_2=\tilde{W}^0, 
\tilde{W}_3=\tilde{H}^0.
\eeq 
Further in Ref.\cite{kos} we identify $L=1$ and $R=2$ in their Eq.(6), 
and we need to complete their Table 1 since the term 
$g(\mu_{L} \tilde{H}^{0} s^{R}_{\mu})$ is missing in Table 1 and one
needs it to expand out Eq.(6). Here we find that the entry for 
the magnitude
for this  coupling in their Table 1 should be the same as the 
magnitude for the coupling  
$g(\mu_{R} \tilde{H}^{0} s^{L}_{\mu})$ listed in Table 1
(see Eqs. (5.1) and (5.4) of Ref.\cite{gunion}).
 Using the above correspondence we find that our result for $2a^{11}_{\mu}$ 
 gotten from our Eq.(25) produces exactly Eq.(6) of Ref.\cite{kos} in the
 limit of vanishing CP phases.

\newpage
\begin{figure}
\begin{center}
\includegraphics[angle=0,width=3.5in]{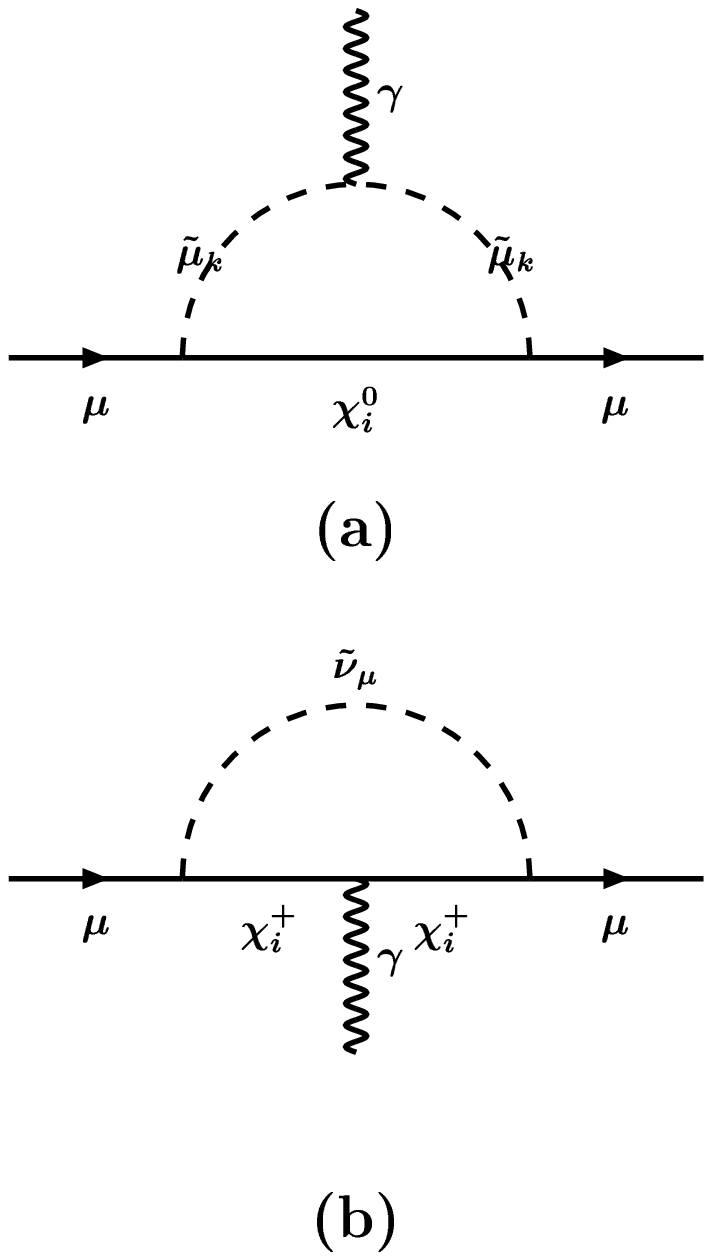}
\caption{The one loop contribution to $g_{\mu}-2$  from 
(a) neutralino exchange, and (b) chargino exchange diagrams.}
\label{fig1}
\end{center}
\end{figure}

\begin{figure}
\begin{center}
\includegraphics[angle=90,width=3.5in]{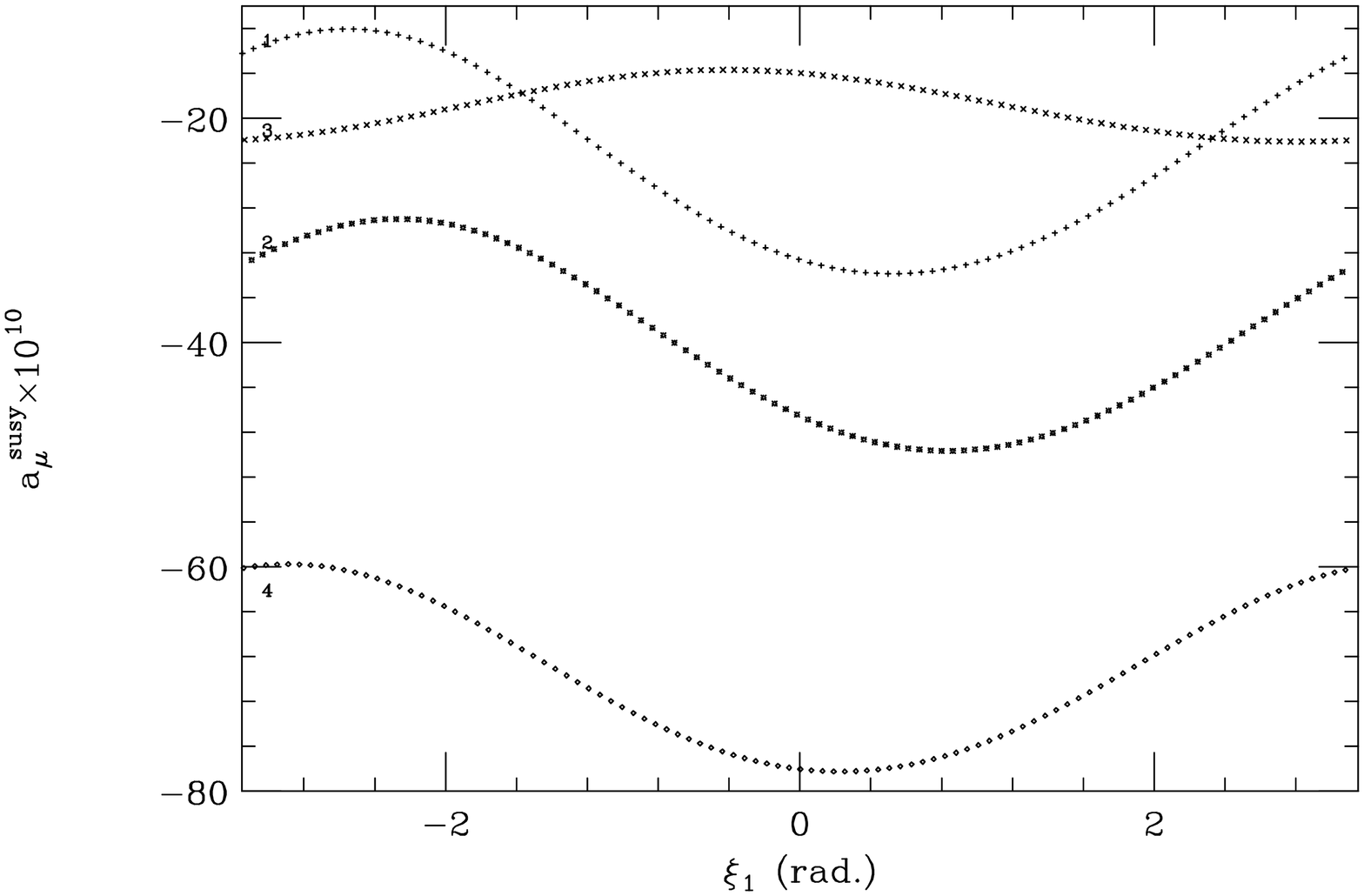}
\caption{Plot of $a_{\mu}^{SUSY}$ as a function of $\xi_1$ without
the imposition of EDM constraints. The values of the other
parameters for the curves (1)-(4) correspond the cases
(1)-(4) in Table 1.  }
\label{fig2}
\end{center}
\end{figure}

\begin{figure}
\begin{center}
\includegraphics[angle=90,width=3.5in]{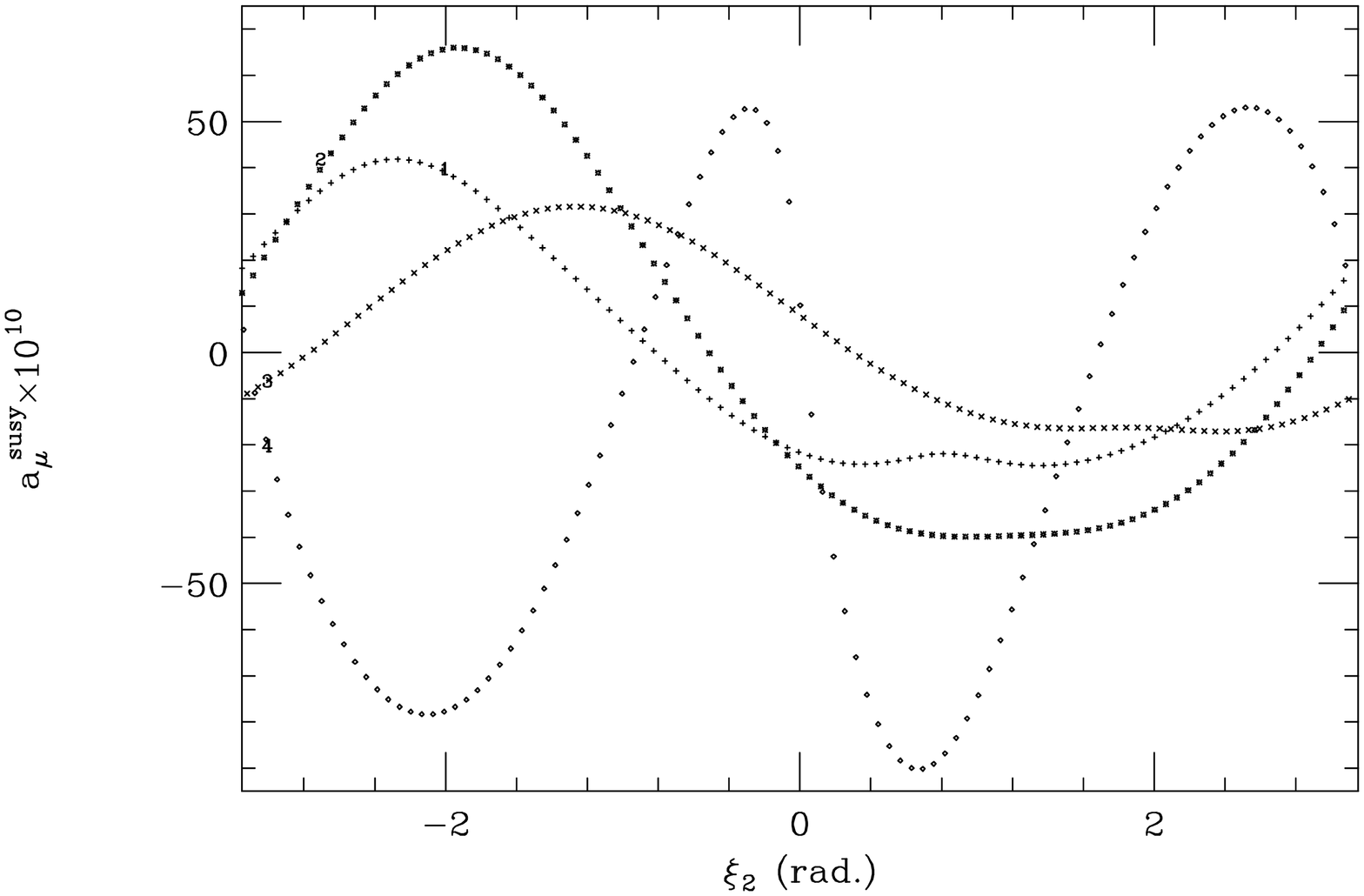}
\caption{Plot of $a_{\mu}^{SUSY}$ as a function of $\xi_2$ without
the imposition of EDM constraints.The values of the other
parameters for the curves (1)-(4) correspond the cases
(1)-(4) in Table 1.  }
\label{fig3}
\end{center}
\end{figure}

\begin{figure}
\begin{center}
\includegraphics[angle=90,width=3.5in]{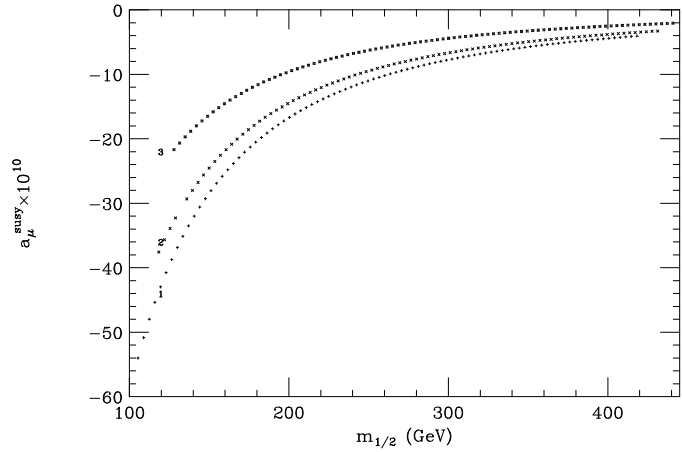}
\caption{Plot of  $a_{\mu}^{SUSY}$ as a function of 
$m_{\frac{1}{2}}$ where all points 
on the trajectories satisfy
the current experimental constraints on the neutron and on the
electron EDM. The curves labelled (1)-(3) are drawn for the 
following set of data: 
(1): $|A_0|$=6.5, $\theta_{\mu}$=2.92, $\alpha_{A_0}$=$-0.4$, $\tan\beta=4$,
$\xi_1=0$, $\xi_2=0.2$, $\xi_3=0.065$;
(2): $|A_0|$=5.4, $\theta_{\mu}$=   3.006, $\alpha_{A_0}$=$-0.1$, $\tan\beta=3.5  $,
$\xi_1=0.105$, $\xi_2=0.105 $, $\xi_3= 0.15 $;
(3): $|A_0|$=2.9   , $\theta_{\mu}$=3.02   , $\alpha_{A_0}$=0.5 , 
$\tan\beta=2.6$,
$\xi_1=0.19$, $\xi_2=0.19$, $\xi_3=0.41$. For all cases $50 
<m_0<250$ (GeV) and all phases are in rad.  }
\label{fig4}
\end{center}
\end{figure}

\end{document}